\documentclass[aps,prb,twocolumn,superscriptaddress]{revtex4}
\usepackage{graphicx}
\usepackage{mathrsfs}
\usepackage{bm}
\usepackage{amsmath}
\usepackage{dcolumn}
\usepackage{epstopdf}
\usepackage{dsfont}
\usepackage{amssymb}
\usepackage{tabularx}
\usepackage{array}
\usepackage{colordvi}
\usepackage{txfonts}
\usepackage{color}
\usepackage{float}

\begin{document}
\title{Berry-Curvature Exchange Induced Anderson Localization in Large-Chern-Number Quantum Anomalous Hall Effect}
\author{Hui Yang}
\affiliation{ICQD, Hefei National Laboratory for Physical Sciences at Microscale, and Synergetic Innovation Center of Quantum Information and Quantum Physics, University of Science and Technology of China, Hefei, Anhui 230026, China.}
\affiliation{CAS Key Laboratory of Strongly-Coupled Quantum Matter Physics, and Department of Physics, University of Science and Technology of China, Hefei, Anhui 230026, China.}
\author{Junjie Zeng}
\affiliation{ICQD, Hefei National Laboratory for Physical Sciences at Microscale, and Synergetic Innovation Center of Quantum Information and Quantum Physics, University of Science and Technology of China, Hefei, Anhui 230026, China.}
\affiliation{CAS Key Laboratory of Strongly-Coupled Quantum Matter Physics, and Department of Physics, University of Science and Technology of China, Hefei, Anhui 230026, China.}
\author{Yulei Han}
\email[Correspondence author:~~]{hanyulei@ustc.edu.cn}
\affiliation{ICQD, Hefei National Laboratory for Physical Sciences at Microscale, and Synergetic Innovation Center of Quantum Information and Quantum Physics, University of Science and Technology of China, Hefei, Anhui 230026, China.}
\affiliation{CAS Key Laboratory of Strongly-Coupled Quantum Matter Physics, and Department of Physics, University of Science and Technology of China, Hefei, Anhui 230026, China.}
\author{Zhenhua Qiao}
\email[Correspondence author:~~]{qiao@ustc.edu.cn}
\affiliation{ICQD, Hefei National Laboratory for Physical Sciences at Microscale, and Synergetic Innovation Center of Quantum Information and Quantum Physics, University of Science and Technology of China, Hefei, Anhui 230026, China.}
\affiliation{CAS Key Laboratory of Strongly-Coupled Quantum Matter Physics, and Department of Physics, University of Science and Technology of China, Hefei, Anhui 230026, China.}

\begin{abstract}
  We theoretically investigate the localization mechanism of quantum anomalous Hall Effect (QAHE) with large Chern numbers $\mathcal{C}$ in bilayer graphene and magnetic topological insulator thin films, by applying either nonmagnetic or spin-flip (magnetic) disorders. We show that, in the presence of nonmagnetic disorders, the QAHEs in both two systems become Anderson insulating as expected when the disorder strength is large enough. However, in the presence of spin-flip disorders, the localization mechanisms in these two host materials are completely distinct. For the ferromagnetic bilayer graphene with Rashba spin-orbit coupling, the QAHE with $\mathcal{C}=4$ firstly enters a Berry-curvature mediated metallic phase, and then becomes localized to be Anderson insulator along with the increasing of disorder strength. While in magnetic topological insulator thin films, the QAHE with $\mathcal{C=N}$ firstly enters a Berry-curvature mediated metallic phase, then transitions to another QAHE with ${\mathcal{C}}={\mathcal{N}}-1$ along with the increasing of disorder strength, and is finally localized to the Anderson insulator after ${\mathcal{N}}-1$ cycling between the QAHE and metallic phases. For the unusual findings in the latter system, by analyzing the Berry curvature evolution, it is known that the phase transitions originate from the exchange of Berry curvature carried by conduction (valence) bands. At the end, we provide a phenomenological picture related to the topological charges to help understand the underlying physical origins of the two different phase transition mechanisms.
\end{abstract}

\maketitle

\section{Introduction}
Anderson localization is one of the most striking transport phenomena in condensed matter physics~\cite{Anderson}. In the presence of disorders, electronic wave function is localized for large enough disorder strength. It is well-known that two-dimensional electrons can be suddenly driven into the Anderson insulating phase even in the presence of extremely weak disorders. When the time-reversal symmetry is broken by applying magnetic fields, or the spin-rotational symmetry is broken by the presence of spin-orbit coupling, a metallic phase appears when the disorder strength $W$ is less than the critical disorder strength $W_{\rm C}$ where the metal-insulator transition occurs~\cite{TRS-broken1,TRS-broken2,SRS-broken}.

Ever since the experimental realization of graphene~\cite{graphene1,graphene2,graphene3} and topological insulators~\cite{topological-insulator1,topological-insulator2}, numerous attention has been paid to explore the long-sought quantum anomalous Hall Effect (QAHE)~\cite{QAHE1,QAHE2,QAHE3,QAHE4,QAHE5,QAHE6,QAHE7,QAHE8,QAHE9,QAHE10,QAHE11,QAHE12,QAHE13,QAHE14,QAHE15,QAHE16,QAHE17,QAHE18,QAHE19,QAHE20,QAHE21,QAHE22,QAHE23,QAHE24,QAHE25,QAHE26,QAHE27,QAHE28}, which has been first experimentally observed in magnetic topological insulator thin films~\cite{XueQiKun} and is still a hot topic on how to increase the observation temperature~\cite{HighTempQAHE}. As a sharp contrast, it is interesting to know how the QAHE will be localized when large external disorders are present. In our previous work~\cite{AndersonQAHE}, we found that the QAHE with Chern number of ${\mathcal{C}}=1$ becomes to be localized into the Anderson insulators by following two completely distinct ways (i.e., QAHE$\rightarrow$Anderson insulator, and QAHE$\rightarrow$ Metal$\rightarrow$Anderson insulator) in the presence of nonmagnetic and spin-flip disorders, respectively. Following the further exploration on how to observe the large-Chern-number QAHE, a natural question arises: how will the large-Chern-number QAHE will be localized in the presence of disorders?

In this article, we investigate the electronic transport properties of the chiral edge modes of large-Chern-number QAHE systems (i.e., bilayer graphene and magnetic topological insulator thin films) in the presence of nonmagnetic and spin-flip disorders. In the ferromagnetic bilayer graphene system with Rashba spin-orbit coupling, the QAHE with Chern number of $\mathcal{C}=4$ can be formed. And in magnetic topological insulator thin films, the QAHE with varying Chern numbers can be produced by tuning the ferromagnetism strength (film thickness) at fixed film thickness (ferromagnetism strength). In the presence of nonmagnetic disorders, we show that both systems become Anderson insulators from the QAHE phase without entering another QAHE phase with different Chern numbers in the presence of strong disorder strength. While for the case of spin-flip disorders, we find that (i) in the bilayer graphene system the QAHE with Chern number of $\mathcal{C}=4$ first enters a Berry-curvature mediated metallic phase and then becomes Anderson insulator along with further increasing the disorder strength; (ii) in the magnetic topological insulator thin film system the QAHE with Chern number of ${\mathcal{C}}=N$ first enters a metallic phase, then enters another QAHE phase with ${\mathcal{C}}={\mathcal{N}}-1$, and is finally fully localized to be Anderson insulator after ${\mathcal{N}} -1$ cycling between the QAHE and metallic phases. In the end, we provide a phenomenological picture to understand the underlying physical mechanisms of the two distinct disorder induced Anderson localization.

\section{System Models and Methods}
In our consideration, we choose the AB-stacked bilayer graphene and magnetic topological insulator thin films to realize the large-Chern-number QAHE. The tight-binding Hamiltonian of the ferromagnetic bilayer graphene in the presence of Rashba spin-orbit coupling can be expressed as~\cite{bilayer graphene1,bilayer graphene2}:
\begin{eqnarray}
H_{\rm BLG}&=&H^{T}_{\rm SLG}+H^{B}_{\rm SLG}+t_{\perp}\sum_{i\in T,j\in B,\alpha}c^{\dag}_{i\alpha}c_{j\alpha},
\label{eq1}\\
H_{\rm SLG}&=&-t\sum_{\langle ij\rangle\alpha}c^{\dag}_{i\alpha}c_{j\alpha}+{\rm i}t_{\rm R}\sum_{\langle ij\rangle\alpha\beta}(s_{\alpha\beta}\times \hat{d}_{ij})\cdot\hat{z}c^{\dag}_{i\alpha}c_{j\beta}\nonumber\\
&+&M\sum_{i\alpha}c^{\dag}_{i\alpha}{s_{z}} c_{i\alpha},
\label{eq2}
\end{eqnarray}
where $H_{\rm SLG}^{T,B}$ represents the Hamiltonian of top (T) or bottom (B) graphene layer. $t_{\perp}$ is the interlayer hopping amplitude. In Eq.~(\ref{eq2}), the first term describes the nearest-neighbor hopping; the second term represents the Rashba spin-orbit coupling with a coupling strength of $t_{\rm R}$, and $\hat{d}_{ij}$ is the unit vector pointing from $j$-site to $i$-site; and the third term is the exchange field with a field strength of $M$. Here, $\alpha$ and $\beta$ denote spin indices.

The tight-binding Hamiltonian of the magnetic topological insulator thin films can be written as~\cite{thin films1}:
\begin{eqnarray}
  H&=&\sum_{i}\varepsilon_{0} c^{\dag}_{i}c_{i}+\sum_{i}\sum_{\gamma=x,y,z} c^{\dag}_{i}T_{\gamma}c_{i+\gamma}+h.c. \nonumber \\
  &+&\sum_{i}m_{0}c^{\dag}_{i}c_{i},
  \label{eq3}
\end{eqnarray}
where $\varepsilon_{0}=(C-{3B}/{2})\tau_{z}\otimes\sigma_{0}$,
$T_{\gamma}={B}/{4}\tau_{z}\otimes\sigma_{0}-{{\rm i}A}/{2}\tau_{x}\otimes\sigma_{\gamma}$, and $m_{0}=m\tau_{0}\otimes\sigma_{z}$. $c_{i}=[a_{i\uparrow},a_{i\downarrow},b_{i\uparrow},b_{i\downarrow}]$, with $(a,b)$ denoting two independent orbitals and $(\uparrow\downarrow)$ representing spin indices. $A, ~B, ~C$ are independent parameters, where $A$ represents the Fermi velocity and C determines the amplitude of the inverted band gap. $\gamma$ is the unit vector towards the direction of $\gamma=(x,y,z)$. $\sigma$ and $\tau$ are respectively spin and orbital Pauli matrices. $m$ measures the effective exchange field strength. Since we focus on the resulting topological phenomena of magnetic thin films, the thickness (i.e., layer number) along $z$-direction is set to be finite~\cite{thin films2}.

In both cases, we apply either nonmagnetic or spin-flip disorders to investigate the transport properties of the large-Chern-number QAHE. The tight-binding Hamiltonian of the applied disorders can be expressed as:
\begin{eqnarray}
H_{\rm dis}=\sum_{i}w^{0}_{i} c^{\dag}_{i}c_{i}+w^{x}_{i}c^{\dag}_{i}\sigma_{x}c_{i}
+w^{y}_{i}c^{\dag}_{i}\sigma_{y}c_{i},
\label{eq3}
\end{eqnarray}
where the first term describes the on-site nonmagnetic disorder, and the last two terms correspond to spin-flip disorders. $w_{0,x,y} $ are uniformly distributed in the interval of $[-W/2,W/2]$ with $W$ representing the disorder strength. In our study, we employ the Landauer-B\"{u}ttiker formula to calculate the two-terminal conductance of the large-Chern-number QAHE in the presence of either nonmagnetic or spin-flip disorders. The disorders are only considered in the central scattering region that couples with two semi-infinite leads, which are exactly extended from the central part. The conductance from right- to left-terminal can be evaluated as~\cite{Data}:
\begin{equation}
G_{\rm LR}=\frac{2e^2}{h}{\rm Tr}\left[\Gamma_{\rm L}G^{\rm r}\Gamma_{\rm R}G^{\rm a}\right],
\label{eq4}
\end{equation}
where $G^{\rm r,a}$ are respectively the retarded and advanced green function, $\Gamma_{\rm L,R}$ are the line-width function coupling with the left and right terminals.

\section{results and discussions}
\begin{figure}
  \includegraphics[width=8.5cm,angle=0]{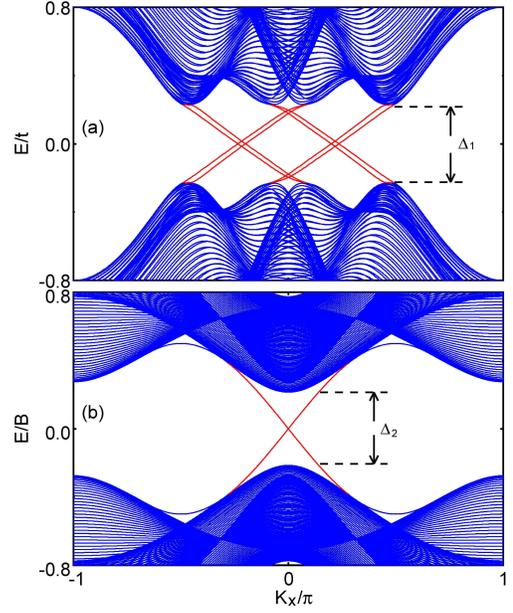}
  \caption{The band structures of nanoribbons for (a) bilayer graphene with Chern number $\mathcal{C}=4$, and (b) magnetic topological insulator thin films with Chern number $\mathcal{C}=2$. Red lines describe the gapless chiral edge modes. $\Delta_{1}$ and  $\Delta_{2}$ denote the band gaps. The corresponding system parameters are set to be $t_{\rm R}=0.4t$ and $M=0.4t$ in panel (a), and $C=0.3B$, $A=0.5B$, $m=1.75B$ in panel (b).}
  \label{fig1}
\end{figure}

Figure~\ref{fig1} displays the band structures of nanoribbons for bilayer graphene with Chern number $\mathcal{C}=4$ and magnetic topological insulator thin films with Chern number $\mathcal{C}=2$, respectively. The widths of the system are respectively set to be $N=60$ and $N=80$. Chiral gapless edge modes highlighted in red lines arise within the energy gap of $\Delta_{1}/t\in [-0.22,0.22] $and $\Delta_{2}/B\in [-0.21,0.21]$, respectively.

\subsection{Localization mechanisms of bilayer graphene based QAHE}
\begin{figure}
  \includegraphics[width=8.5cm,angle=0]{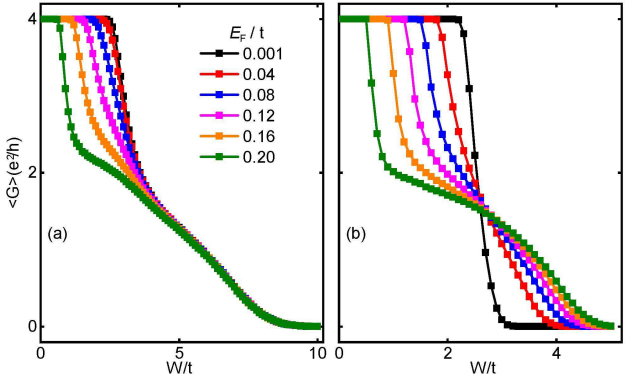}
  \caption{Averaged conductance $\langle G\rangle$ as a function of $W$ for different Fermi energies inside the bulk gap. The system is bilayer graphene with parameters set to be $N=60$, $t_{\rm R}=0.4t$ and $M=0.4t$. (a) and (b): For nonmagnetic and spin-flip disorders, respectively. 4000 samples are collected for each data point. }
  \label{fig2}
\end{figure}

In the bilayer graphene based QAHE system, we select several representative Fermi energies inside the bulk band gap, i.e., $E_{\rm F}/t=0.001,~0.04,~0.08,~0.12,~0.16,~0.20$ to investigate the disorder effect on the electronic transport properties. Figure~\ref{fig2} depicts the averaged two-terminal conductance $\langle G\rangle$ as a function of the disorder strength $W$ for respective nonmagnetic and spin-flip disorders. For the nonmagnetic case as displayed in Fig.~\ref{fig2}(a), the conductance keeps quantized to be $\langle G\rangle=4e^{2}/h$ in the presence of weak disorders. When the disorder strength reaches certain critical strength, the conductance suddenly becomes to decrease first and then gradually decrease to be vanishing. In particular, at a fixed disorder strength, the closer of the Fermi energy near the charge neutrality point $E_{\rm F}/t=0.00$, the more robust of the averaged conductance against disorder.
However, for the spin-flip disorder case, the disorder effect on the electronic transport properties exhibit completely different from those observed in the nonmagnetic case. Figure~\ref{fig2}(b) shows the averaged conductance as a function of disorder strength in the presence of spin-flip disorder. One can find that the variation of the averaged conductance is divided into two parts with the increase of disorder strength, i.e. (i) in the presence of weak disorder $(W/t \textless 2.3)$, the averaged conductance keeps quantized ($\langle G\rangle=4~e^2/h$) before it reaches the critical disorder strength, (ii) after disorder exceeds the critical strength, the averaged conductance gradually vanishes. In contrast to the nonmagnetic case, the farther of the Fermi energy away from the charge neutrality point $E_{\rm F}/t=0.00$, the more difficult of the conductance to be destroyed. For instance, the averaged conductance for $E_{\rm F}/t=0.001$ keeps quantized until $(W/t \approx 2.2)$ and once $W/t>2.3$, it immediately drops to zero.

\subsection{Localization mechanism for $\mathds{Z}_2$ topological insulator thin film based QAHE}
\begin{figure}
  \includegraphics[width=8.5cm,angle=0]{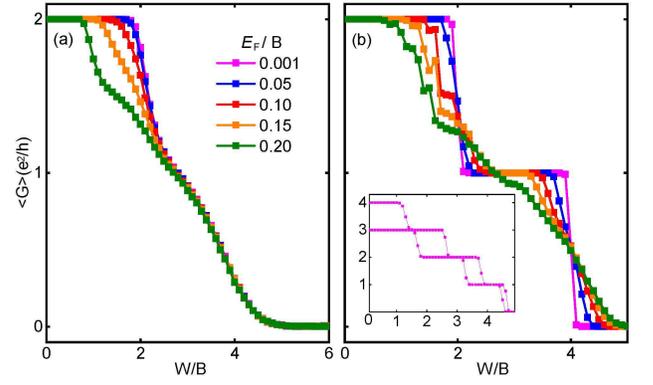}
  \caption{Averaged conductance $\langle G\rangle$ as a function of $W$ for different Fermi energies inside the bulk gap. The system is magnetic topological insulator thin flims with width set to be $N=80$. (a) and (b): For nonmagnetic and spin-flip disorders, respectively. 4000 samples are collected for each data point. Inset of (b) is the conductance variation of system with higher Chern number in $E_{\rm F}/B$=0.001. 300 samples are collected for each data point. The parameters in the inset are $C=0.3B$, $A=0.5B$, $m=1.71B$, $N=60$ for $\sigma_{xy}=4e^{2}/h$ and $C=0.3B$, $A=0.5B$, $m=1.7B$, $N=60$ for $\sigma_{xy}=3e^{2}/h$, respectively.}
  \label{fig3}
\end{figure}

In the magnetic topological insulator thin films, Fig.~\ref{fig3} displays the averaged conductance as a function of disorder strength for different Fermi energies inside the bulk gap. In the presence of nonmagnetic disorder [see Fig.~\ref{fig3}(a)], the variation of averaged conductance with the increase of disorder strength is similar with the nonmagnetic case of bilayer graphene system [see Fig.~\ref{fig2}(a)]. While in the presence of spin-flip disorders, as shown in Fig.~\ref{fig3}(b), the system exhibits a completely different localization phenomenon in contrast to the above numerical results. The most striking feature is the appearance of additional quantized conductance plateau as the increases of disorder strength $W$, i.e., in the presence of weak disorder, the averaged conductance keeps quantized with a value of $\langle G\rangle=2e^{2}/h$; when $W$ reaches the first critical value (e.g., $W_{\rm C1}/B \approx 1.90$ at $E_{\rm F}/B=0.001$), the initial quantized plateau is destroyed and the system quickly enters a lower quantized plateau with a value of $\langle G\rangle=e^{2}/h$; as $W$ continuing to increase, the averaged conductance keeps quantized at $\langle G\rangle=e^{2}/h$; when $W$ reaches the second critical point (e.g., $W_{\rm C2}/B \approx 3.90$ at $E_{\rm F}/B=0.001$), the quantized plateau is destroyed and the conductance gradually vanishes. Note that, the closer of the Fermi energy to $E_{\rm F}/B=0.00$, the more robust of the lower quantized plateau against disorder, therefore the lower quantized plateau almost disappears at the conduction band edge ($E_{\rm F}/B=0.20$).

For magnetic topological insulator thin films, the Chern number can be tuned by varying the thickness of the films. Inset of Fig.~\ref{fig3}(b) shows the averaged conductance as a function of disorder strength for the system of ${\mathcal{C}}=4$ and ${\mathcal{C}}=3$ at $E_{\rm F}/B=0.001$. One can see that the Hall conductance is destroyed step by step and forms plateaus in every integer value. To understand the fundamental physics that results in these anomalous transport properties, we analyze the Berry curvature density in the corresponding bulk system that reflects the nature of the QAHE.

\begin{figure}[htp!]
	\includegraphics[width=8.5cm,angle=0]{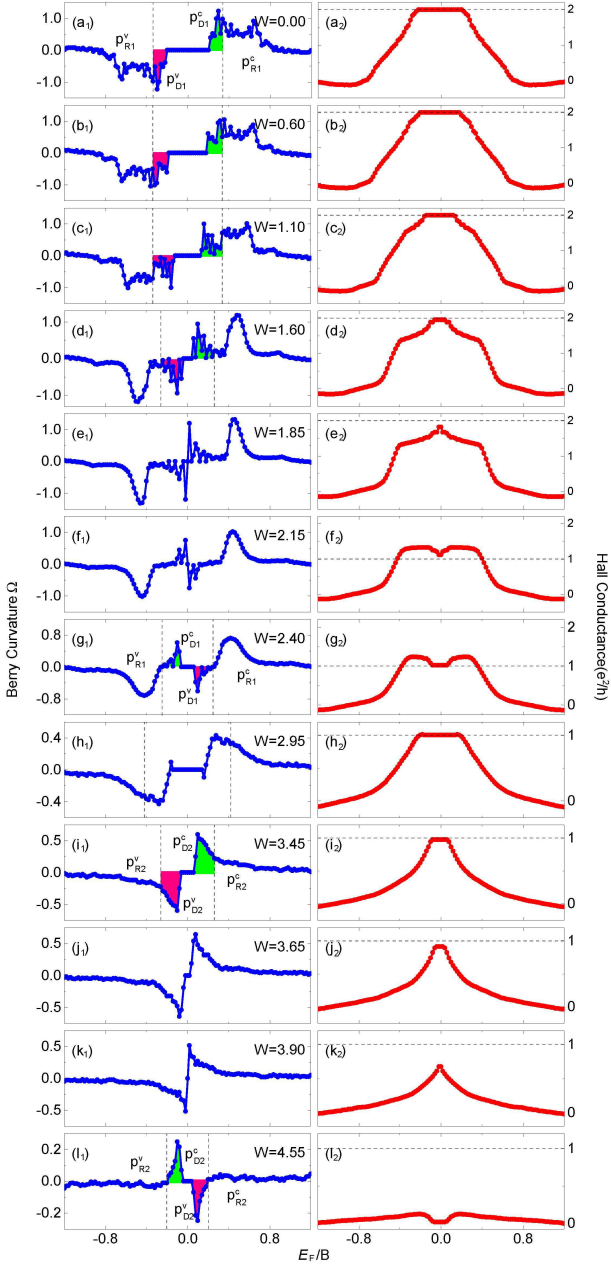}
	\caption{$\rm (a_{1})$-$\rm (l_{1})$ Evolution of averaged Berry curvature density $\Omega$ as a function of energy $E$ for different  spin-flip disorder strengths $W$. $\rm (a_{2})$-$\rm (l_{2})$ Corresponding averaged Hall conductance $\sigma_{xy}$ as a function of energy $E$. $\rm P^{v/c}_{D1/2}$ is the conductance of first/second Dirac point. v/c represents valence/conduction bands. $\rm P^{v/c}_{R1/2}$ is the conductance of remaining bands. 30 samples are collected for each point.}
	\label{fig4}
\end{figure}

\begin{figure}[htp!]
	\includegraphics[width=8.5cm,angle=0]{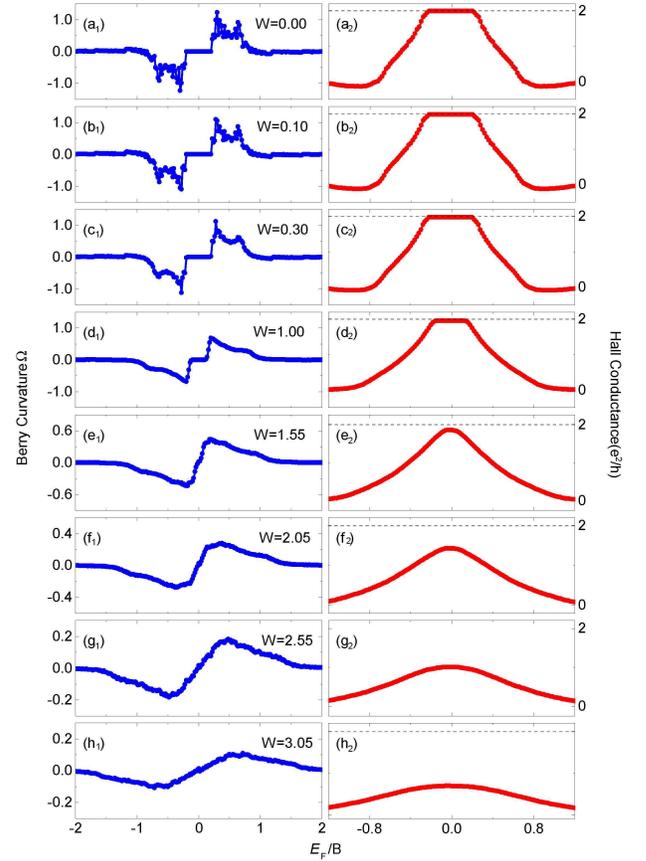}	
	\caption{$\rm (a_{1})$-$\rm (l_{1})$ Evolution of averaged Berry curvature density $\Omega$ as a function of energy $E$ for different  nonmagnetic disorder strengths $W$. $\rm (a_{2})$-$\rm (l_{2})$ Corresponding averaged Hall conductance $\sigma_{xy}$ as a function of energy $E$. 30 samples are collected for each point.}
	\label{fig5}
\end{figure}

\begin{figure*}[htbp]
  \centering
  \includegraphics[width=17.2cm,angle=0]{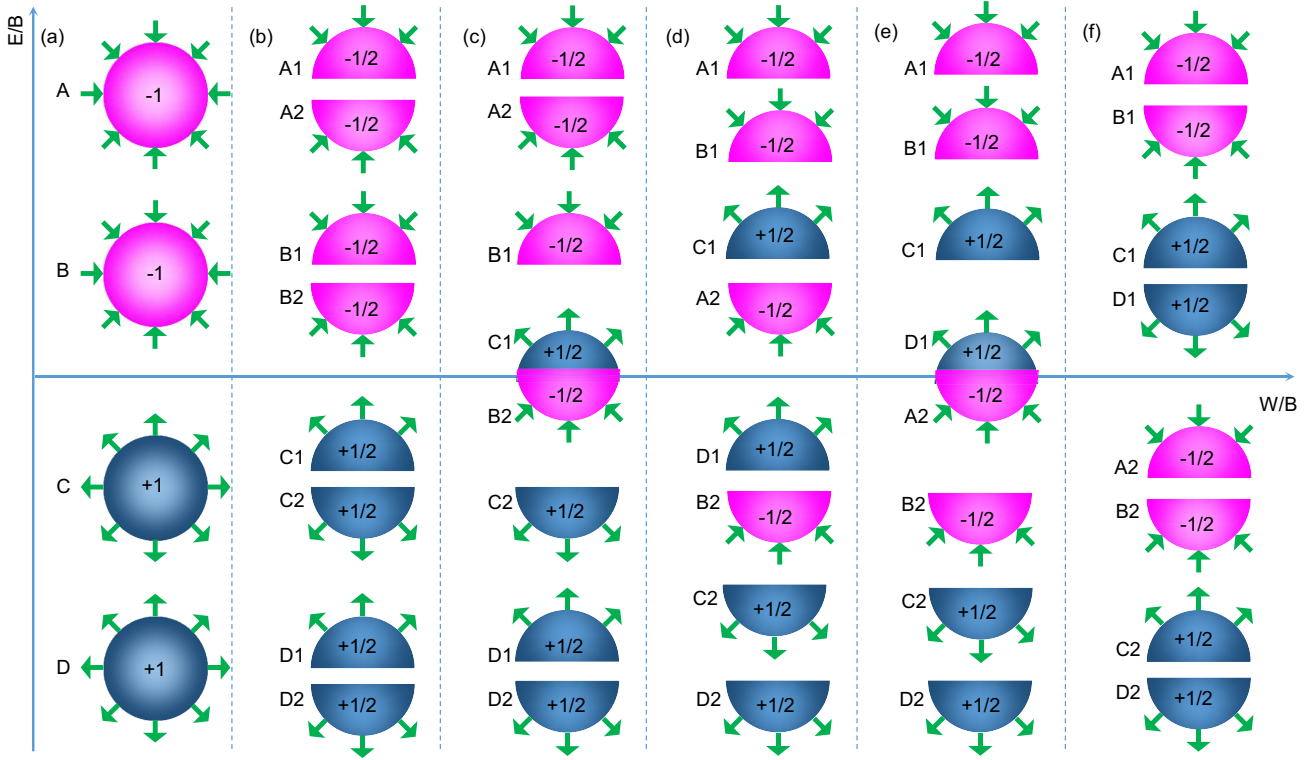}	
  \caption{Schematic of the evolution of topological charges carried by the spin-textures of the valence and conduction bands in magnetic topological insulator thin films with spin-flip disorders. (a) In the absence of disorders, valence/conduction bands carry four Skyrmions with topological charges of +1 +1 -1 -1. (b) At weak disorders, there four Skyrmions are scattered to be eight Merons that carry half-integer topological charges and are labelled as ``A1",``A2", ``B1", ``B2", ``C1", ``C2", ``D1" and ``D2". (c) By further increasing disorder strength, Merons B2 and C1 move towards each other. (d) Merons B2 and C1 make an exchange leading to Hall conductance falling on next ladder. (e),(f) Weak disorders hardly affect on Merons A2 and D1. When disorder strength becomes to the second critical point, Merons A2 and D1 move towards each other and make a exchange result in the disappearance of the integer conductance.}
  \label{fig6}
\end{figure*}

\begin{figure*}[htbp]
  \centering
  \includegraphics[width=17.2cm,angle=0]{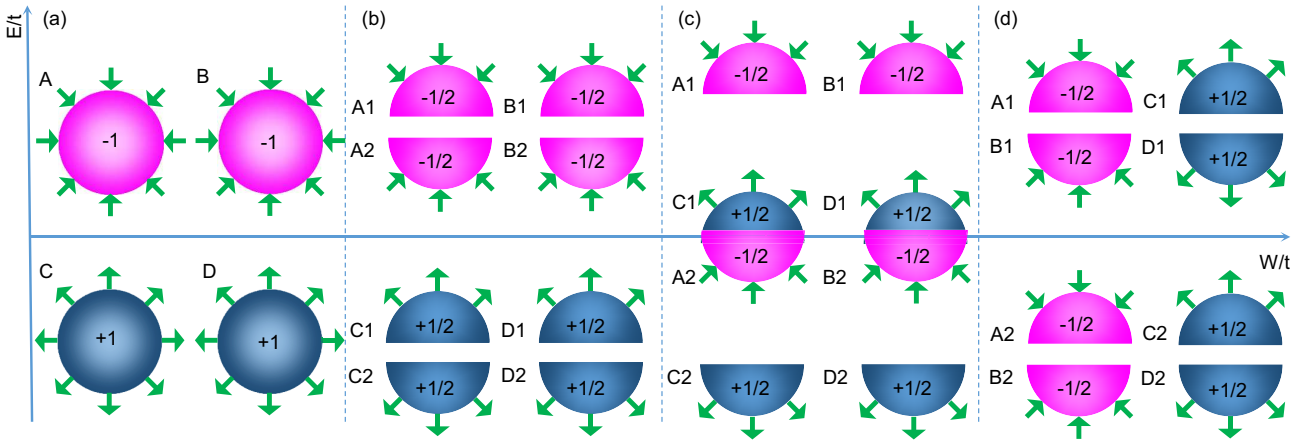}	
  \caption{Schematic of the evolution of topological charges carried by the spin-textures of the valence and conduction bands in bilayer graphene with spin-flip disorders. (a) In the absence of disorders , valence/conduction bands carry four Skyrmions with topological charges of +1 +1 -1 -1. (b) At weak disorders, there four Skyrmions are scattered to be eight Merons that carry half-integer topological charges and are labelled as ``A1", ``A2", ``B1", ``B2", ``C1", ``C2", ``D1" and ``D2". (c) By further increasing disorder strength, Merons A2 and C1, B2 and D1 move towards each other. (d) At strong disorders, Merons A2 and C1, B2 and D1 exchange together causing the Anderson localization.}
  \label{fig7}
\end{figure*}

\subsection{Berry curvature analysis}
For the disordered system, we use a generalized Berry curvature and Kubo formula to calculate conductances in real space instead of reciprocal space which can be expressed as\cite{Berryanalysis}
 \begin{equation}
 \Omega_{\alpha}=-\sum\limits_{\beta\not=\alpha}\frac{{\rm 2Im}\langle\alpha|v_{x}|\beta\rangle\langle\beta|v_{y}|\alpha\rangle}{(w_{\alpha}-w_{\beta})^{2}}
 \end{equation}
 \begin{equation}
 \sigma_{xy}=-\frac{e^{2}}{h}\int\langle\Omega(\epsilon)\rangle f(\epsilon)\rm d\epsilon
 \end{equation}
In the above equations, $\Omega(\epsilon)=\frac{1}{ S}{\rm Tr}[\hat{\Omega}\delta(\epsilon-H)]$ describes the Berry curvature density in the energy spectrum with $\hat{\Omega}$ being the Berry curvature operator $\hat{\Omega}_{\alpha}=\sum_{\alpha}|\alpha\rangle\langle\alpha|\ $ where $|\alpha\rangle$ indicates the eigenvector of $|\hbar w_{\alpha}\rangle$ in the disorderd system. $S$ is the area of the two-dimensional system and $ v_{x/y}$ is the velocity operator in the $x$ or $y$ direction. $f(\epsilon)$ is the Fermi-Dirac distribution.

In the bilayer graphene, the transport properties of nonmagnetic or spin-flip disorder doped systems are almost same as that in the square lattice with Chern number ${\mathcal{C}}=1$ and monolayer graphene with Chern number ${\mathcal{C}}=2$, which have been studied in our previous paper~\cite{QAHE20}. Therefore, in this article, we mainly explore the mechanism of anomalous transport properties in the magnetic topological insulator thin films.

In the magnetic topological insulator thin films, Fig.~\ref{fig4} shows the evolution of the averaged Berry curvature density $\Omega$ and Hall conductance $\sigma_{xy}$ as a function of the Fermi energy $E$ for different spin-flip disorders strengths $W/B=0.00, 0.60, 1.10, 1.60, 1.85, 2.15, 2.40, 2.95, 3.45, 3.65, 3.90,$ $4.25, 4.55$. At $W/B=0.00$ [see Fig.~\ref{fig4}(a$_1$)], we can identify two Berry curvature peaks in either valence bands or conduction bands as $\rm P_{D1}^{v,c}$ that from the contribution of massive Dirac bands, and $\rm P_{R1}^{v,c}$ that from the contribution of remaining bands, where $\rm v/c$ denotes valence/conduction bands and $\rm D1/R1$ denotes the first Dirac/remaining bands.
One can find that for Fermi energies inside the bulk band gap, $\Omega(E)=0$ and $\sigma_{xy}=2e^{2}/{h}$, and the Hall conductance preserve $\sigma_{xy}(E)=\sigma_{xy}(-E)$ due to the electron-hole symmetry of energy bands.
In the presence of weak disorders [see Figs.~\ref{fig4}(b)$-$\ref{fig4}(d)], the two Berry curvature peaks $\rm P_{D1}^{v,c}$ move towards each other and shrink the band gap. The contribution of areas covered by red/green color to Hall conductance are about $0.5e^{2}/{h}$. Meanwhile, the two Berry curvature peaks $\rm P_{R1}^{v,c}$ are almost fixed and broaden. When the disorder strength approaches the first critical point $W/B\approx 1.85$ [Fig.~\ref{fig4}(e)], the bulk gap begins closing and the Berry curvature peaks $\rm P_{R1}^{v,c}$ start to exchange with each other. If the disorder strength continues to increase [see Figs.~\ref{fig4}(f)-(h)], one can find that the two peaks $\rm P_{R1}^{v,c}$ make a interchange with the bulk gap reopening, and the Hall conductance inside the bulk gap now becomes $\sigma_{xy}=e^{2}/{h}$, which is consistent with above transport calculations. At even stronger disorders [see Fig.~\ref{fig4}(i)], one can see that the two peaks $\rm P_{R1}^{v,c}$ vanish and instead two new peaks appear, which can be identified as $\rm P_{D2}^{v,c}$ followed by the similar definitions above. When the disorder strength is strong [see Figs.~\ref{fig4}(j)-(l)], the two Dirac peaks $\rm P_{D2}^{v,c}$ undergo the second interchange process, and finally the Hall conductance inside the bulk gap is zero, indicating the system becomes the Anderson insulator.

In contrast, Fig.~\ref{fig5} displays how the averaged Berry curvature density and Hall conductance evolve for the magnetic topological insulator thin films with nonmagnetic disorders. With the increase of disorder strength, although the two Dirac peaks move towards each other shrinking the bulk gap, the most noticeable difference from the case of spin-flip disorders is the lack of interchange of Berry curvature carried by two Dirac peaks. Therefore, the system with nonmagnetic disorders does not undergo a Berry-curvature mediated metallic phase but transit directly from the QAHE state to the Anderson insulating phase.

\subsection{A phenomenological picture}
From above observed anomalous transport phenomenon in large Chern number system, we can now provide a phenomenological picture of topological charge to understand it.

In the magnetic topological insulator thin films, the continuous model of half Bernevig–Hughes–Zhang model is $H=k_{x}\sigma_{x}+k_{y}\sigma_{y}+m\sigma_{z}$. Owing to infinite domain of integration instead of first Brillouin zone in reciprocal space, the Hall conductance carried by a Dirac point is about $0.5e^{2}/h$. Before band inversion, the Hall conductance carried by Dirac point and remaining bands is respectively $0.5e^{2}/h$ and $-0.5e^{2}/h$ in valence bands, thus the total Hall conductance carried in valence bands is zero. After band inversion, the Hamiltonian is transformed to $H=k_{x}\sigma_{x}+k_{y}\sigma_{y}-m\sigma_{z}$. Now the Hall conductance carried by Dirac point and remaining bands is respectively $-0.5e^{2}/h$ in valence bands, so we could realize $-e^{2}/h$ QAHE. Here the disorder strength plays a role of a mass term in aforementioned band inversion mechanism. From the topological point of view, the quantized Hall conductances are analogous to four Skyrmions, as shown in Fig. ~\ref{fig6}(a), where valence/conduction bands carry four topological charges of +1 +1 -1 -1 labelled as ``A'', ``B", ``C'', ``D", and spins pointing outwards (inwards) is +1 (-1). When the spin-flip disorders are applied [see Fig.~\ref{fig6}(b)], all Skyrmions will be divided to eight merons (labelled as "A1", "A2", ``B1", ``B2", ``C1", ``C2", ``D1" and ``D2"). When the disorder strength increasing to the first critical point [see Fig.~\ref{fig6}(c)], merons B2 and C1 move towards each other and begin to interchange.
At stronger disorder strength [see Fig.~\ref{fig6}(d)], merons B2 and C1 have finished the interchange, thus the Hall conductance in valence bands is transit to the lower plateau $\sigma_{xy}= e^{2}/{h}$. Note that, merons D1 and A2 would not be influenced by the spin-flip disorders until disorder strength approaches the second critical point. When the disorder strength continuing to increase [see Fig~\ref{fig6}(e)-(f)], merons D1 and A2 undergo the interchange process, thus the total Hall conductance vanishes.

In the bilayer graphene system, the mechanism of Berry curvature interchange is different from that in magnetic topological insulator thin films. The key point is that for all four Skyrmions are totally equivalent. Fig.~\ref{fig7} displays the evolution of topological charges carried by the spin-textures of the valence and conduction bands in bilayer graphene system. In the presence of weak disorder [see Fig.~\ref{fig7}(a)-(b)], four Skyrmions (for simplicity, we draw four Skyrmions. Actually, there are eight Skyrmions in bilayer graphene because of the Chern number $\mathcal{C}=4$) are scattered to eight merons labelled as ``A1", ``A2", ``B1", ``B2", ``C1", ``C2", ``D1" and ``D2". At stronger disorder [see Fig.~\ref{fig7}(c)], the merons A2 and C1, B2 and D1 not make an interchange step by step, but interchange at the same time. After the interchange [see Fig.~\ref{fig7}(d)], the net charge in valence band is zero, indicating the system enters the Anderson insulating phase.

As aforementioned discussion, we know that the mechanisms of the phase transition are different in the bilayer graphene and magnetic topological insulator thin films based QAHE system. The reason is that in the bilayer graphene system, the Skyrmions induced by nonzero Chern number are equivalent leading to the undistinguished feature during the phase transition, making the system enters the Anderson insulator after once interchange. Whereas in the magnetic topological insulator thin films, repeated band inversion will produce multiple unequal Skymions and the destruction condition of these Skymions are disparate by spin- flip disorders.

\section{Conclusion}
In summary, we theoretically investigate disorders effect (including nonmagnetic disorder and spin-flip disorder) on the two types of QAHE systems with large Chern number. In nonmagnetic disorder case, both of the systems display similar localization behavior that directly transit from the QAHE state to the Anderson insulating state. In spin-flip disorder case, we find that with the increase disorder strength, both of the systems exist the Berry curvature mediated metallic phase but the mechanisms of phase transition are different. In magnetic topological insulator thin films, the quantized conductances are destroyed step by step from higher to lower integer plateau and finally the system enters the Anderson insulating phase. Whereas in bilayer graphene, the quantized conductances are destroyed in only one step and eventually the system becomes the Anderson insulator. The underlying physical mechanism in the two types of QAHE systems can be explained from the view of topological charges, where considering valence/conduction bands carry different charges respectively. Our work provide a deep understanding to the disorder effect on the QAHE systems with tunable Chern number.

\begin{acknowledgments}
This work was financially supported by the the National Natural Science Foundation of China (No. 11974327 and No. 11474265), National Key Research and Development Program (No. 2016YFA0301700 and No. 2017YFB0405703), and Anhui Initiative in Quantum Information Technologies. We are grateful to the supercomputing service of AM-HPC and the Supercomputing Center of USTC for providing the high-performance computing resources.
\end{acknowledgments}

\end{document}